\documentclass[aip,rsi,reprint,twocolumn,
amsmath,amssymb,amsfonts,showpacs,showkeys]{revtex4-2}
\usepackage{graphicx}

\begin{document}
	\title{Design and construction of the multiplexing cold neutron spectrometer BOYA with double-column Rowland focusing analyzers}
	
	\author{Jinchen Wang}
	\affiliation{Laboratory for Neutron Scattering, School of Physics, Renmin University of China, Beijing 100872, China}
	\affiliation{Key Laboratory of Quantum State Construction and Manipulation (Ministry of Education), Renmin University of China, Beijing, 100872, China}
	
	\author{Daye Xu}
	\affiliation{Laboratory for Neutron Scattering, School of Physics, Renmin University of China, Beijing 100872, China}
	\affiliation{Key Laboratory of Quantum State Construction and Manipulation (Ministry of Education), Renmin University of China, Beijing, 100872, China}
	
	\author{Juanjuan Liu}
	\affiliation{Laboratory for Neutron Scattering, School of Physics, Renmin University of China, Beijing 100872, China}
	\affiliation{Key Laboratory of Quantum State Construction and Manipulation (Ministry of Education), Renmin University of China, Beijing, 100872, China}
	
	\author{Wei Luo}
	\affiliation{Laboratory for Neutron Scattering, School of Physics, Renmin University of China, Beijing 100872, China}
	\affiliation{Institute of High Energy Physics, Chinese Academy of Sciences
	(CAS), Beijing 100049, China}
	\affiliation{Spallation Neutron Source Science Center, Dongguan 523803, China}
	
	\author{Peng Cheng}
	\affiliation{Laboratory for Neutron Scattering, School of Physics, Renmin University of China, Beijing 100872, China}
	\affiliation{Key Laboratory of Quantum State Construction and Manipulation (Ministry of Education), Renmin University of China, Beijing, 100872, China}
	
	\author{Hongxia Zhang}
	\email[Corresponding author: ]{hxzhang@ruc.edu.cn}
	\affiliation{Laboratory for Neutron Scattering, School of Physics, Renmin University of China, Beijing 100872, China}
	\affiliation{Key Laboratory of Quantum State Construction and Manipulation (Ministry of Education), Renmin University of China, Beijing, 100872, China}
	
	\author{Wei Bao}
	\email[Corresponding author: ]{weibao@cityu.edu.hk}
	\affiliation{Laboratory for Neutron Scattering, School of Physics, Renmin University of China, Beijing 100872, China}
	\affiliation{Department of Physics, City University of Hong Kong, Kowloon, Hong Kong SAR}
	\affiliation{Center for Neutron Scattering, City University of Hong Kong, Kowloon, Hong Kong SAR}
	
	\begin{abstract}
	Developing neutron spectrometers with higher counting efficiency has been an essential pursuit in neutron instrumentation.
	In this work, we present BOYA, a multiplexing cold neutron spectrometer designed and implemented at the China Advanced Research Reactor.
	Equipped with 34 angular analyzing channels spanning 119\textdegree{}, each containing 5 inelastic channels and 1 diffraction channel, BOYA enhances the measurement efficiency by two orders of magnitude over a traditional triple-axis spectrometer.
	To optimize both intensity and energy resolution, innovative double-column Rowland focusing analyzers have been developed. By filling the crystal gaps in the traditional Rowland focusing geometry, our design enhances the neutron beam coverage without introducing appreciable double-scattering. Our commissioning results on vanadium and MnWO$_4$ have confirmed the success of the design, establishing BOYA as a successful multiplexing instrument for neutron spectroscopy.
	\end{abstract}
	
	
	\maketitle

	\section{Introduction}
	Inelastic neutron scattering spectrometers have played significant roles across various research fields. 
	Notably, the triple-axis spectrometer (TAS), which analyzes the momentum and energy transfer of the neutron beam by adjusting instrumental angles \cite{Brockhouse,TripleAxis}, has been a Nobel Prize winning tool for studying elemental excitations in solid state physics, such as the dispersion of magnons and phonons.
	However, TAS detects a single point in the vast four dimensional momentum-energy phase-space at each angular configuration.
	Mapping out an excitation spectrum on TAS is a lengthy process.
	As neutron flux is still quite limited even at the most advanced neutron sources, and the demand on spectrometer time continues to grow, developing spectrometers with higher counting efficiency and broader phase-space coverage has become a major theme in the community.
	
	There are two approaches to increase the coverage in neutron spectrometers. 
	One approach involves using pulsed neutrons, allowing the analysis of neutron energies by the time-of-flight (TOF) of neutron events at an array of detectors \cite{TOF93}. 
	Recent examples of direct geometry TOF spectrometers include LET \cite{LET} at ISIS, CNCS and HYSPEC \cite{CNCS1,HYSPEC,comparison} at Oak Ridge National Laboratory (ORNL), and AMATERAS \cite{AMATERAS} at the Japan Proton Accelerator Research Complex (J-PARC).
	The other approach, particularly suited to steady-state neutron sources, is multiplexing. 
	This technique increases the number of analyzer-detector channels in TAS, enabling simultaneous measurement of multiple final momentum and energy values ($\boldsymbol{k}_f$,$E_f$).
	Early multiplexing instruments, keeping the analyzers scattering plane horizontal, increased the number of inelastic channels by up to one order of magnitude \cite{RITA_1,RITA_II_2,book_Igor}.
	Further expansion was difficult due to space constraints between analyzing channels.
	MACS \cite{MACS96,MACS} at NIST Center for Neutron Research (NCNR) made an exceptional breakthrough to accommodate 20 inelastic channels by adopting double-crystal analyzers. 
	
	By rotating the analyzer scattering plane from horizontal to vertical, a denser coverage is possible, as exemplified by the FlatCone at ILL \cite{Flatcone}.
	More recently, further advances have been made by the CAMEA (Continuous Angle Multiple Energy Analysis) configuration \cite{CAMEA_concept}, which installs a series of vertically scattered energy analyzers in each angular channel along with quasi-continuous angular coverage in the horizontal plane. 
	It has greatly increased the number of inelastic channels and has stimulated the developments of multiplexing instruments, such as CAMEA \cite{CAMEA_PSI} at PSI, BIFROST \cite{BIFROST1,BIFROST2} at the European Spallation Source, MultiFLEXX \cite{MultiFLEXX2,MultiFLEXX3} at Helmholtz-Zentrum Berlin and BAMBUS \cite{BAMBUS} at FRM II.

	In this work, we report on BOYA (meaning `broad and elegant' in Chinese), a novel multiplexing cold neutron spectrometer at the China Advanced Research Reactor (CARR), Beijing. 
	Innovative double-column Rowland focusing (DCRF) analyzers were designed for BOYA, to optimize the intensity and energy resolution at the backend.
	By introducing a second column of highly oriented pyrolytic graphite (HOPG) crystals and optimizing the sample-to-analyzer and analyzer-to-detector distances, this design fills the crystal gaps in the traditional Rowland focusing analyzer, and enhances the coverage of the scattered beam without significant double-scattering. Commissioning experiments on the incoherent elastic scattering of vanadium and the spin wave excitation of MnWO$_4$ demonstrate the success of this multiplexing instrument.

	\section{Design}
	
	\begin{figure}[tbhp]
		\centering
		\includegraphics[width=\columnwidth]{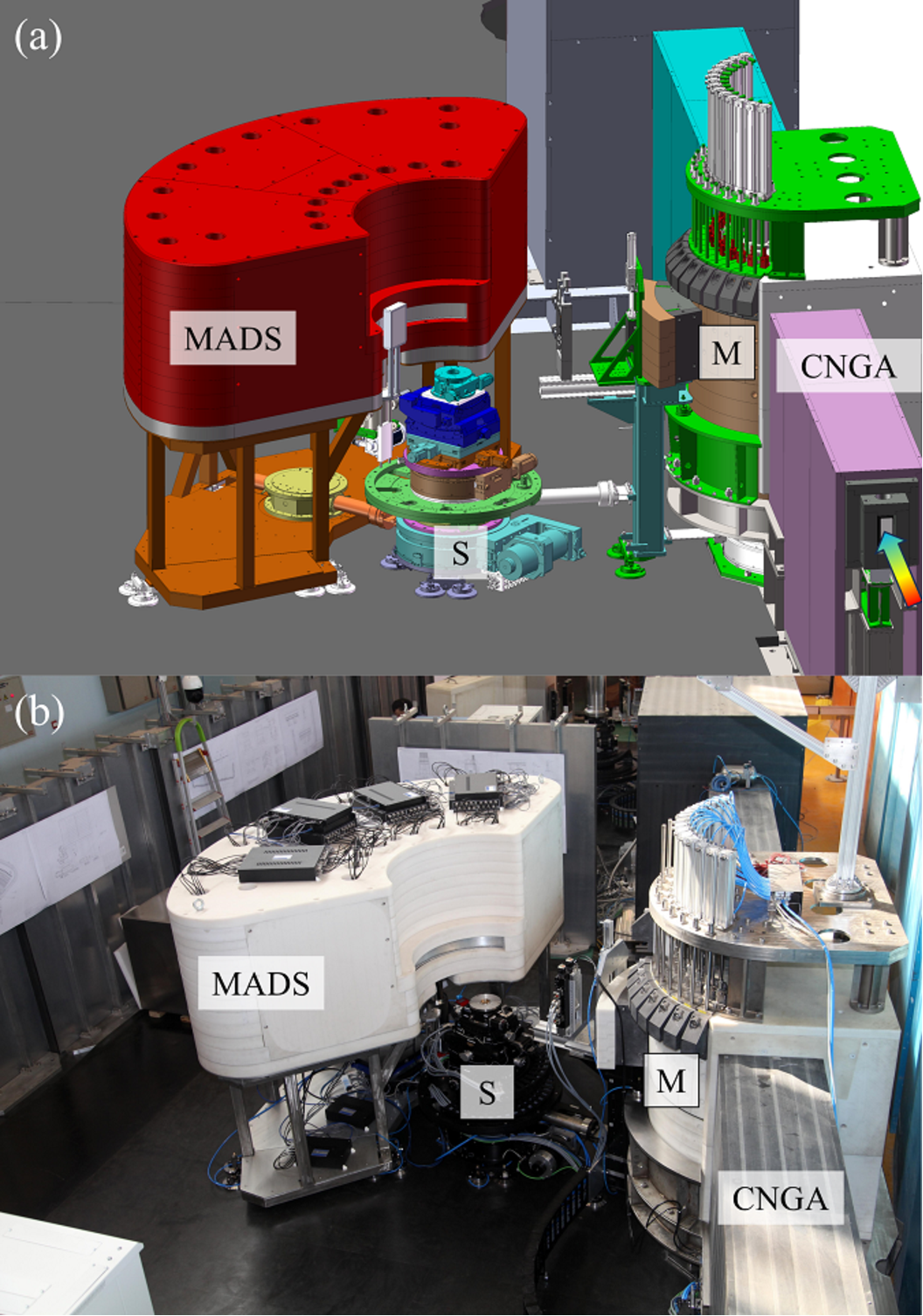}
		\caption{(a) Design diagram and (b) actual photograph of the BOYA instrument. The CARR cold neutron guide A (CNGA), monochromator (M) with its liftable shielding structure, sample stage (S) and the multi-analyzer-detector-system (MADS) are indicated. The direction of the neutron beam in the guide is marked with a colored arrow in (a).
		}
		\label{overview}
	\end{figure}
	
	\begin{figure}[tbhp]
		\centering
		\includegraphics[width=\columnwidth]{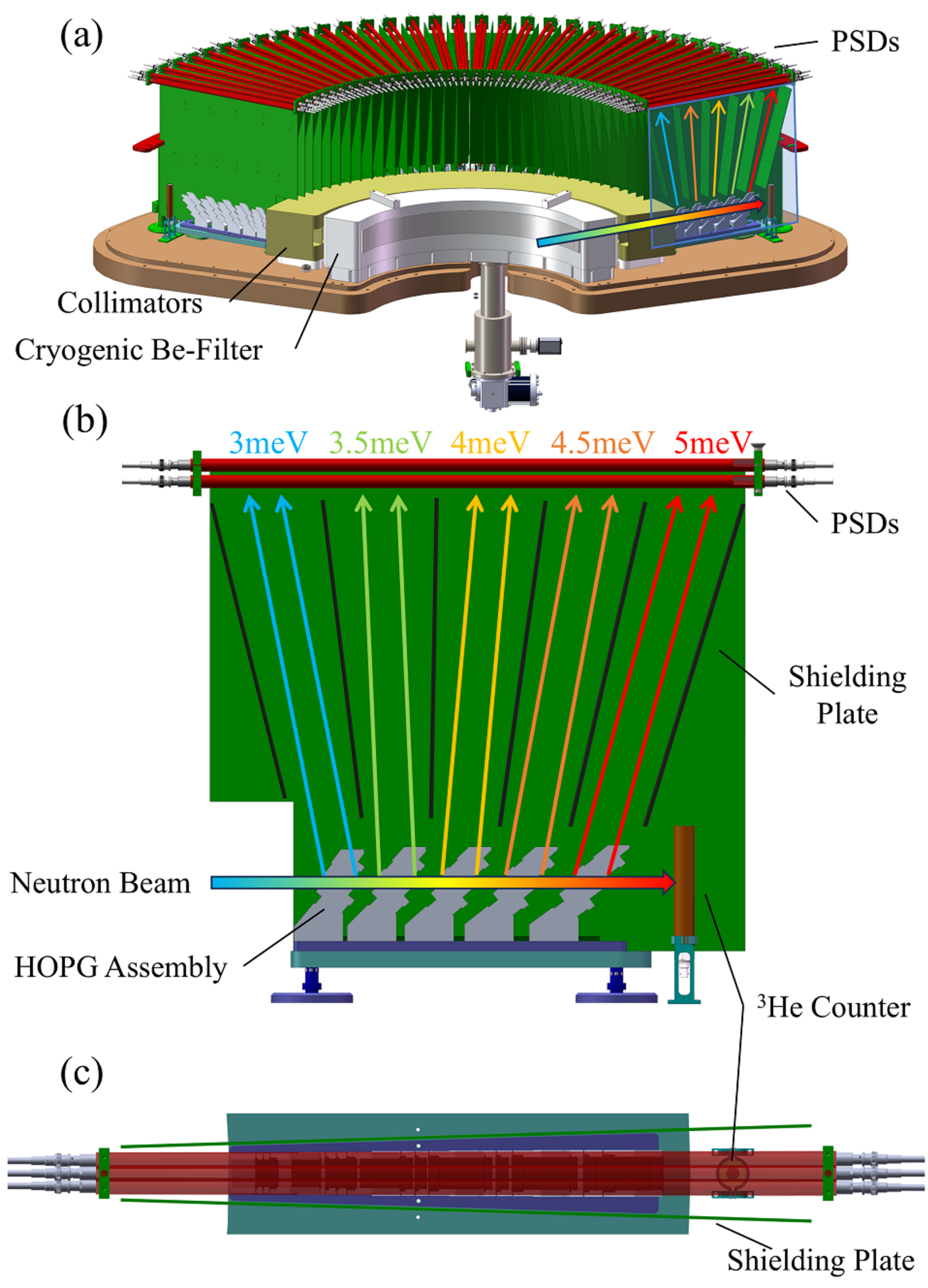}
		\caption{ (a) Design diagram inside the MADS tank, spanning 119\textdegree{} horizontal angle. The cryogenic Be-filter, collimators, position sensitive detectors (PSDs) are indicated. One angular channel is highlighted by a blue box. 
		(b) Side view of the highlighted angular channel. 
		The neutron beam is represented by a rainbow-colored arrow. The analyzers, composed of HOPG assemblies, scatter neutrons upwards and produce two focused peaks for each final energy. The shielding plates made of borated aluminum are indicated.
		(c) Top view of a group of three PSDs covering one angular channel.
		}
		\label{MADS}
	\end{figure}

	BOYA is stationed at the first beam port of the CARR cold neutron guide A (CNGA), which has an m-value of 2.
	Consisting of a 17\,m long curved guide with a radius of curvature of 1253\,m and an inner cross section of 30\,mm $\times$ 150\,mm, the guide filters out fast neutrons below the critical wavelength of 2 $\AA$.
	As shown in Fig.~\ref{overview}, the spectrometer includes the monochromator, sample stage, and the multi-analyzer-detector system (MADS) which is integrated and shielded in a 100\,mm thick borated polyethylene (BPE, 5\%wt B$_2$O$_3$) tank.
	The MADS is air-floated and movable on the dancing floor, providing tunable scattering angle coverage depending on the incident energy. 
	The monochromator shares the same design with XINGZHI \cite{XINGZHI,zhang_PG_2024}, a cold neutron TAS located at the end of the CNGA guide, adjacent to BOYA. 
	Consisting of 3 columns and 7 rows of HOPG crystals with an effective surface of 105\,mm (wide) $\times$ 154\,mm (high), the monochromator can deliver a vertically focused beam with incident energy ($E_i$) ranging from 2.5\,meV to 15\,meV. 
	The HOPG crystals (provided by Panasonic) used in the monochromator and in analyzers described below, have X-ray characterized mosaic spreads within the range of 0.4\textdegree{}--0.5\textdegree{}.
	The monochromator is shielded by a B$_4$C/Pb/BPE sandwich structure with pneumatic liftable shielding blocks \cite{Wang_Shielding}.

	The inner structure of MADS is displayed in Fig.~\ref{MADS}.
	The 34-channeled cryogenic Beryllium (Be) filter and collimators are placed at the entrance of the MADS.
	In the cryogenic chamber of the Be-filter, each channel houses a Be block of 28\,mm (wide) $\times$ 100\,mm (high) $\times$ 120\,mm (long, in the neutron flight path), shielded by borated (31\% B$_4$C) aluminum plates on surrounding sides (top, bottom, left, and right). 
	Curved aluminum windows of 0.5\,mm thick are welded on the front and back sides of the entire Be-filter chamber. 

	\begin{table}[tbp]
	\caption{Specifications of BOYA.}
	\label{BOYA_Specs}
	\begin{ruledtabular}
		\begin{tabular}{ll}
			Guide Cross Section	& 30\,mm $\times$ 150\,mm \\
			Monochromator	& PG(002), vertical focusing \\
			Incident Energy & 2.5 to 15\,meV \\
			Focused Beam Size &  30\,mm $\times$ 30\,mm\\
			M-to-S Distance \footnote{Monochromator-to-sample distance.}	
			& 1.6-2.1\,m \\
			Beam Height from Ground & 1.3\,m \\
			Scattering Angle & 
			15\textdegree{}--116\textdegree{} at $E_i = 5$\,meV \\
			Fixed Final Energies & 3.0, 3.5, 4.0, 4.5, 5.0\,meV \\
			Angular Channels \footnotemark[2]
			& 34 angular channels covers 119\textdegree{} \\
			Total Number of Channels \footnotemark[2]
			& 34$\times$(5+1) \\
		\end{tabular}
	\end{ruledtabular}
	\footnotetext[2]{Currently half of them are installed.} 
	\end{table}

	The 34 angular channels of MADS span 119\textdegree{} in the horizontal plane, with a separation of 3.5\textdegree{} between channels. 
	Within each angular channel, a series of energy analyzers is installed, enabling the simultaneous analysis and detection of five fixed final energies ($E_f$) at 3.0, 3.5, 4.0, 4.5, 5.0\,meV. 
	Neutrons of these corresponding energies are scattered upwards, where a group of three position sensitive detectors (PSDs) is horizontally installed at the top of each angular channel. 
	Each PSD has a diameter of 16\,mm, an effective length of 680.8\,mm and a spatial resolution of 6--8\,mm along the tube. The three PSDs, at heights of 490 and 510\,mm relative to the horizontal beam, are arranged to cover at least 90\% of the reflected beam from the analyzers. 
	The radial arrangement of PSD tubes tolerates minor misalignments of the analyzer crystals, allowing the intensity to be captured even when the analyzer-reflected beam slightly shifts or broadens.
	To optimize signal, the counts from the three-PSD group are typically summed to represent a single scattering angle during inelastic experiments. However, the tubes can be distinguished individually when higher angular resolution is prioritized instead of intensity.
	A $^3$He counter, serving as the diffraction channel, is located to capture the elastically diffracted neutrons at each angular channel.
	Borated aluminum plates are used to prevent cross-talk between angular and energy channels. 
	The key specifications of BOYA are listed in Table~\ref{BOYA_Specs}.
	
	\begin{figure}[tbhp]
	\centering
	\includegraphics[width=\columnwidth]{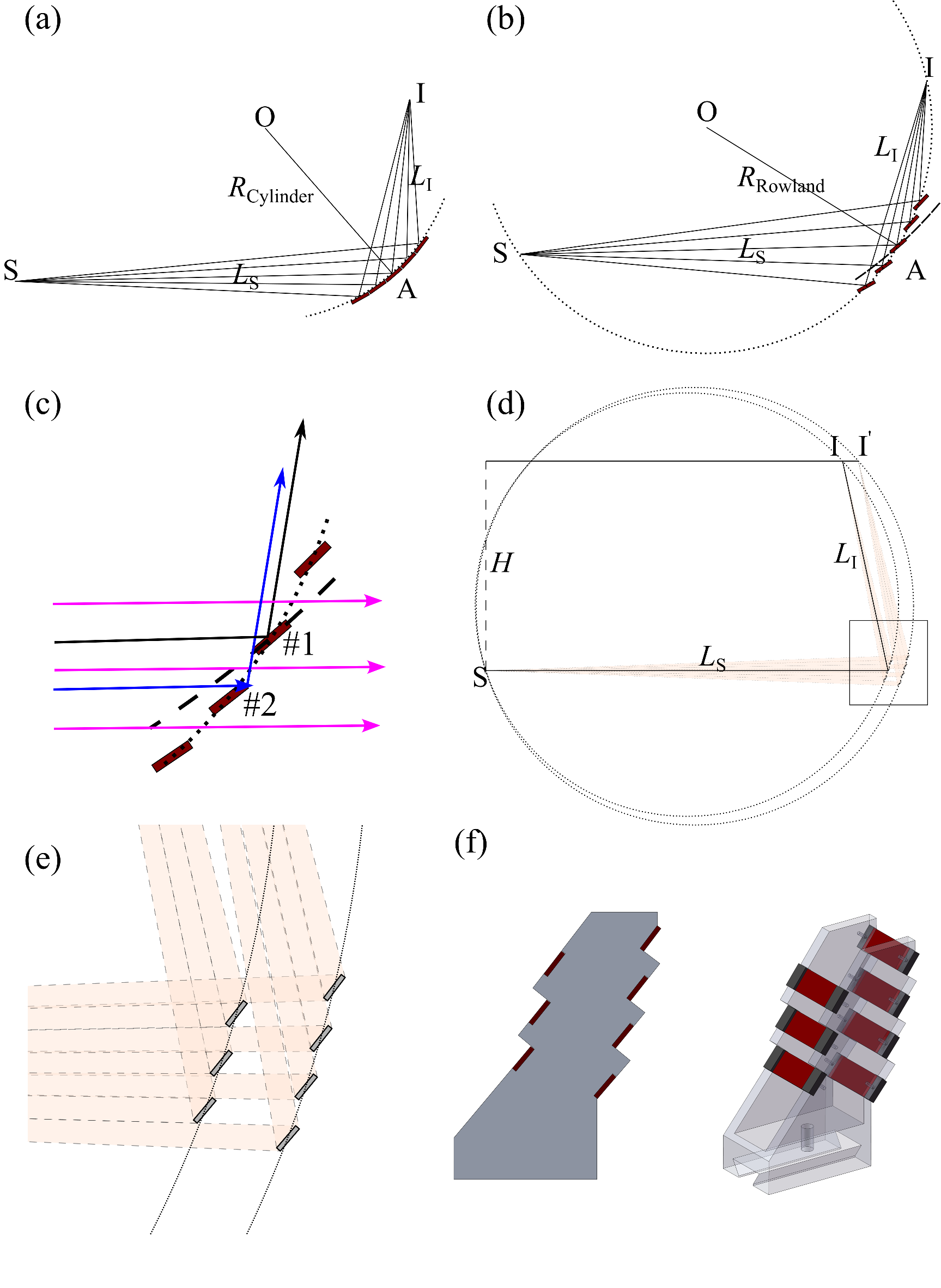}
		\caption{(a) Schematic drawing of cylindrical focusing and (b) Rowland focusing. The source (S), analyzer (A), image (I), center of circle (O), radius of the cylinder ($R_{\mathrm{Cylinder}}$) and Rowland circle ($R_{\mathrm{Rowland}}$) are indicated. The source-to-analyzer and analyzer-to-image distances are denoted by $L_\mathrm{S}$ and $L_\mathrm{I}$.
		(c) Enlarged view of the crystal arrangement for Rowland focusing. The crystal blades are shown as rectangular blocks, two of which labeled as \#1 and \#2. The dotted curve shows the Rowland circle, and the dashed line is what crystal surfaces are parallel to. The black arrow represents the reflected beam by the center of blade \#1, the blue arrow shows the beam connecting the edges of blades \#1 and \#2, and the magenta arrows show the beam passing through the gaps. 
		(d) Schematic drawing of the DCRF analyzer, with two Rowland circles and two images (I and I$^\prime$) indicated. The vertical height of PSDs is denoted as $H$. The box highlights the detailed arrangement of the crystal blades, which is enlarged in (e). In (e), the neutron beam, shaded in pink, is reflected by the complementary first and second columns without double-scattering. 
		(f) Diagram of the DCRF analyzer, with the aluminum support and HOPG blades denoted in grey and red, respectively.
		}
	\label{focus}
	\end{figure}
	
	\begin{figure*}[tbhp]
	\centering
	\includegraphics[width=0.85\textwidth]{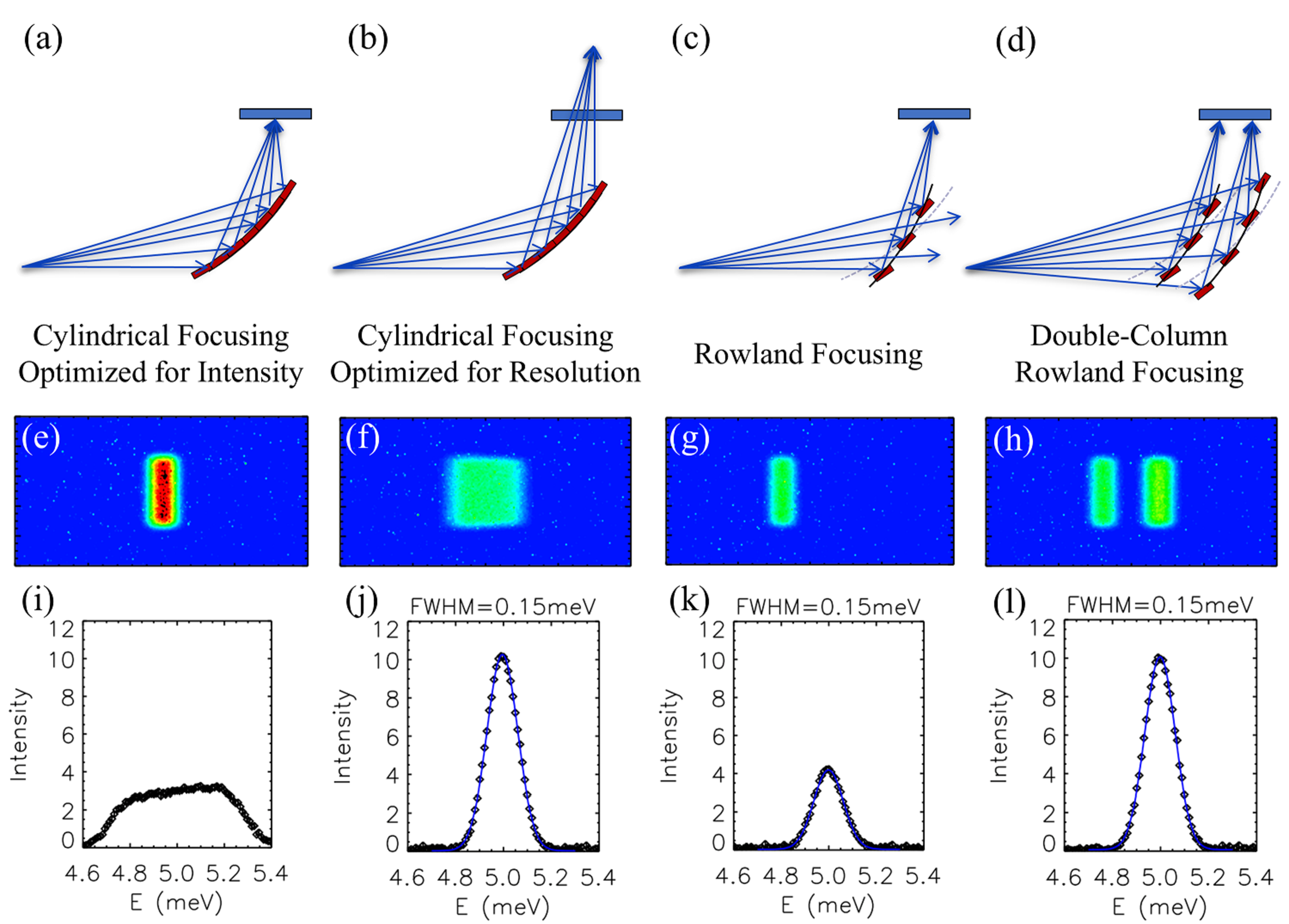}
	\caption{ Simulation results of different focusing schemes, including
		(a) cylindrical focusing optimized for intensity (b) cylindrical focusing optimized for resolution, (c) traditional Rowland focusing and (d) double-column Rowland focusing. The corresponding (e-h) spatial distribution and  (i-l) energy distribution on the area detector of 200\,mm $\times$ 100\,mm are shown. In (j-l), the full width at half maximum (FWHM) from Gaussian peak fits are labeled at the top of plots.
		}
	\label{ana_sim}
	\end{figure*}

	To achieve focusing, there are different options to arrange HOPG crystals in multi-crystal analyzers \cite{book_Igor}. In cylindrical focusing, HOPG blades are tangential to the circumference of a cylinder, with the radius $R_{\mathrm{Cylinder}}$ given by:
	\begin{equation}
		R_{\mathrm{Cylinder}} = \frac{2L_\mathrm{S}L_\mathrm{I}}{(L_\mathrm{S} + L_\mathrm{I})\sin\theta_A}
	\end{equation}
	where $L_\mathrm{S}$ and $L_\mathrm{I}$ are the source (S) to the analyzer (A) and the analyzer (A) to the image (I) distances, respectively, and $2\theta_A$ is the take-off angle of the central analyzer blade, as shown in Fig.~\ref{focus}(a).
	In this arrangement, the crystal surface acts like a curved mirror, focusing the paraxial beam from S to I. In general, the take-off angle $2\theta_A$ changes slightly across the blades, yielding a polychromatic beam at I, except in the symmetrical case when $L_\mathrm{S} = L_\mathrm{I}$. 
	
	Another commonly used focusing scheme, shown in Fig.~\ref{focus}(b), is Rowland focusing, where S, I and the center of each crystal blade lie on the Rowland circle. The radius $R_{\mathrm{Rowland}}$ is given by: 
	\begin{equation}
		R_{\mathrm{Rowland}} = \frac{\sqrt{L_\mathrm{S}^2 + L_\mathrm{I}^2 + 2L_\mathrm{S}L_\mathrm{I}\cos2\theta_A}}{2\sin2\theta_A}
	\end{equation}
	This achieves a monochromatic focusing as the reflecting angle $2\theta_A$ remains constant across the blades.  
	Rowland focusing and its prismatic variants \cite{Prismatic} are employed by multiplexing spectrometers such as CAMEA \cite{CAMEA_concept,CAMEA_PSI}. 
	It is important to note that the crystal surface is not tangential to the Rowland circle. Instead, they are tangential to another circle with a radius $(L_\mathrm{S} + L_\mathrm{I})/2\sin\theta_A$, tilted by an angle from the Rowland circle \cite{MACS96,Rowland_Simul}. This generates a stair-like arrangement of crystal blades.
	If the near-parallel blades are positioned too closely, double-scattering would occur, where neutrons reflected from the front surface of the first blade are further scattered by the back surface of the nearby second blade. This effect was also recognized in the design considerations of MultiFLEXX \cite{MultiFLEXX3}.
	To avoid appreciable double-scattering, gaps are introduced to separate adjacent blades, as shown in Fig.~\ref{focus}(c). While the black arrow represents the expected beam path, the blue arrow connecting the edges of blade \#1 and \#2 highlights the critical condition where a minimum gap is necessary to avoid double-scattering. However, the gap results in significant loss of neutron flux, as illustrated by magenta arrows.

	\begin{figure}[tbhp]
		\centering
		\includegraphics[width=\columnwidth]{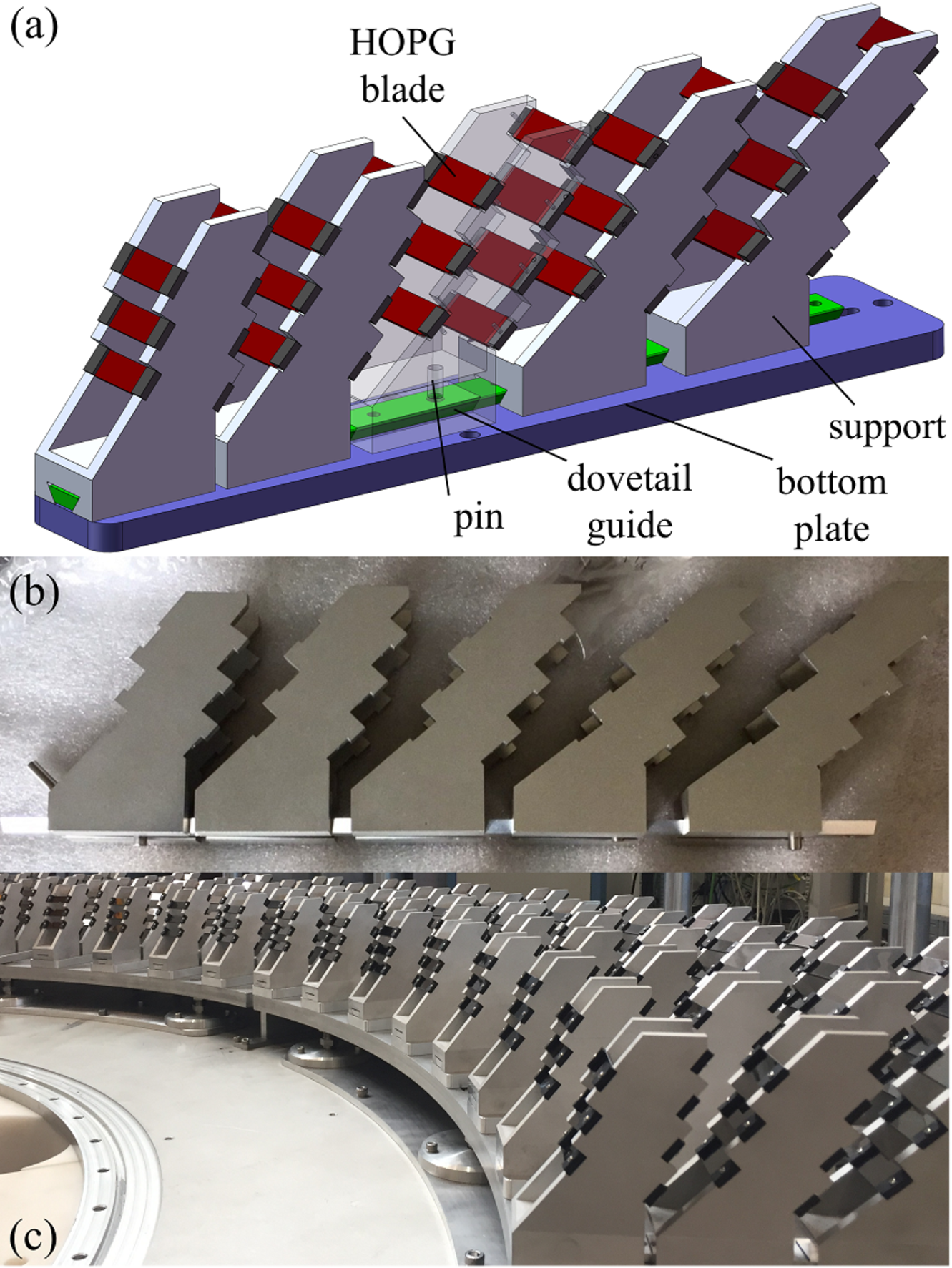}
		\caption{ (a) The design diagram of a series of five DCRF analyzers, with fixed $E_f$ of 3.0, 3.5, 4.0, 4.5, 5.0\,meV. The middle support is displayed with transparency to show the inner structure. (b) Photo of five DCRF analyzers in an angular channel. (c) Photo of radially installed analyzer arrays.
		}
		\label{ana_real}
	\end{figure}

	We have developed the DCRF analyzer, which fills crystal gaps, avoids double-scattering and optimizes the energy resolution at the same time.
	The key innovation is to introduce a second column of HOPG crystals, which are aligned on a second Rowland circle including the source S and the second image I$^\prime$ (see Fig.~\ref{focus}(d)). I$^\prime$ and I are of the same height $H$. The two columns of crystals can complement each other's gaps.
	To achieve this, it is noted that the minimum gap size to avoid double-scattering is determined by $L_\mathrm{S}$ and $L_\mathrm{I}$ values. When the ratio $L_\mathrm{S}/L_\mathrm{I}$ is adjusted to around 2, the minimum gap size matches the size of the crystal blade approximately.
	This allows the crystals in the front and back columns to cover the beam completely, as illustrated in Fig.~\ref{focus}(e). 
	
	To verify the design, we performed ray-tracing Monte Carlo simulations using McStas \cite{McStas}, shown in Fig.~\ref{ana_sim}. 
	A white source with a $10\times10\,\mathrm{mm^2}$ cross section and a uniform energy distribution around 5\,meV was placed at the origin. 
	The beam was reflected by HOPG crystals (0.45\textdegree{} mosaic spread) under different focusing schemes, including cylindrical focusing optimized for intensity, cylindrical focusing optimized for resolution, traditional Rowland focusing, and double-column Rowland focusing, as illustrated in Fig.~\ref{ana_sim}(a-d).
	The asymmetric condition of $L_\mathrm{S} = 1265\,\mathrm{mm}$ and $L_\mathrm{I} = 520\,\mathrm{mm}$ were used, matching BOYA's actual values. An incoherent scatterer (vanadium, not shown) was included to mimic stray background on the detectors. 
	The spatial and energy distributions of the reflected beam were monitored at the detector position.
	The results are shown in Fig.~\ref{ana_sim}(e-l).

	The simulations show that for cylindrical focusing, optimization of intensity and resolution cannot be achieved at the same time. 
	For cylindrical focusing optimized for resolution, the beam is focused onto the detector, but its energy spreads out.
	For cylindrical focusing optimized for resolution, the focus lies at a distance equal to $L_\mathrm{S}$, beyond the detector plane. 
	This results in a spread-out beam on the detector.
	Although the total intensity is conserved, the signal-to-noise ratio is reduced. 
	Traditional Rowland focusing has optimized energy resolution as well as focused beam on the detector. But it suffers from half intensity loss due to gaps between crystal blades, which are inevitable to prevent double-scattering.
	The DCRF produces two image spots on the detector as expected (Fig.~\ref{ana_sim}(h)). 
	It not only optimizes energy resolution but also recovers the intensity loss, doubling the intensity compared to traditional Rowland focusing, as shown in Fig.~\ref{ana_sim}(l). 
	
	Our installed DCRF analyzers are shown in Fig.~\ref{ana_real}.
	The DCRF design is realized by mounting seven HOPG blades on a high precision aluminum support manufactured for each final energy. The crystal blades are shown in red and crystal mounts covered by B$_4$C rubbers shown in black.
	Five supports fit onto the dovetail guide via the dovetail mating groove. To ensure good positioning, one pin is inserted through the support and the guide. Screws are tightened between the bottom plate and the guide so that the guide secures the analyzer support in place (see Fig.~\ref{ana_real} (a-b)). All the bottom plates are finally connected to a big base plate through screws and two pins, forming a radial arrangement shown in Fig.~\ref{ana_real} (c).

	\begin{figure}[tbhp]
		\centering
		\includegraphics[width=\columnwidth]{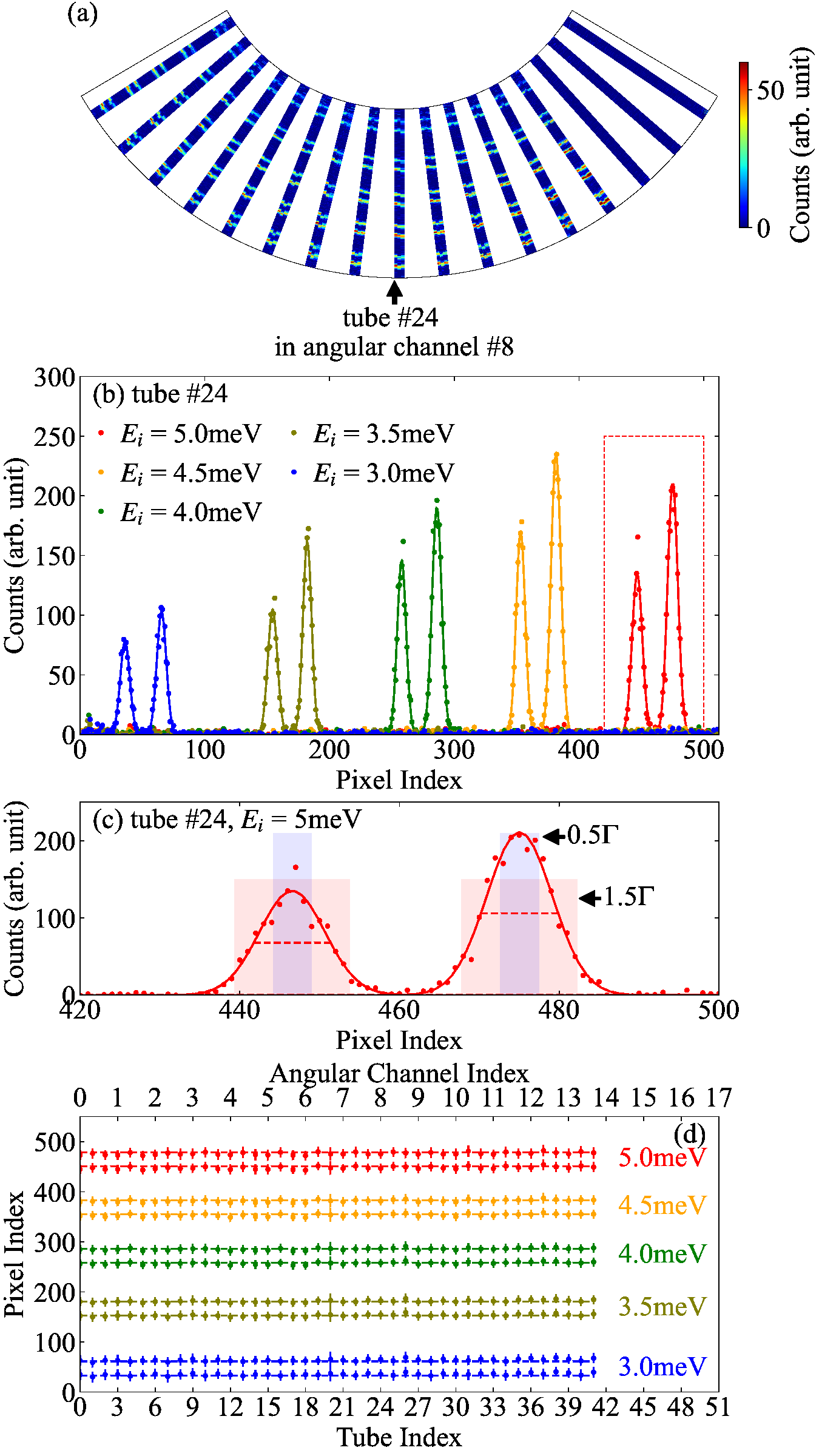}
		\caption{ The focused images from fixed $E_i$ values of 3.0, 3.5, 4.0, 4.5 and 5.0\,meV. (a) Intensities maps across all PSDs distributed on the MADS. Note that the rightmost three channels are blocked by the beam stop. (b) Intensity distribution along the tube \#24, divided into 512 pixels. Smaller pixels correspond to the lower $E_f$ end. Data from different $E_i$ are represented by dots of different colors, and Gaussian-fit curves represented by colored lines.
		The region within the dashed box for $E_i = 5\,\mathrm{meV}$ is enlarged in (c), where the FWHM ($\Gamma$), 0.5\,FWHM ($0.5 \Gamma$), 1.5\,FWHM ($1.5 \Gamma$) of two peaks are marked by the dash lines, the blue shades and the red shades, respectively. 
		(d) Image positions on different PSD tubes, with FWHMs represented by error bars. The calculated positions are indicated by dash lines. The upper-axis labels the angular channel index.
		}
		\label{detview}
	\end{figure}
	
	\begin{figure}[tbhp]
		\centering
		\includegraphics[width=\columnwidth]{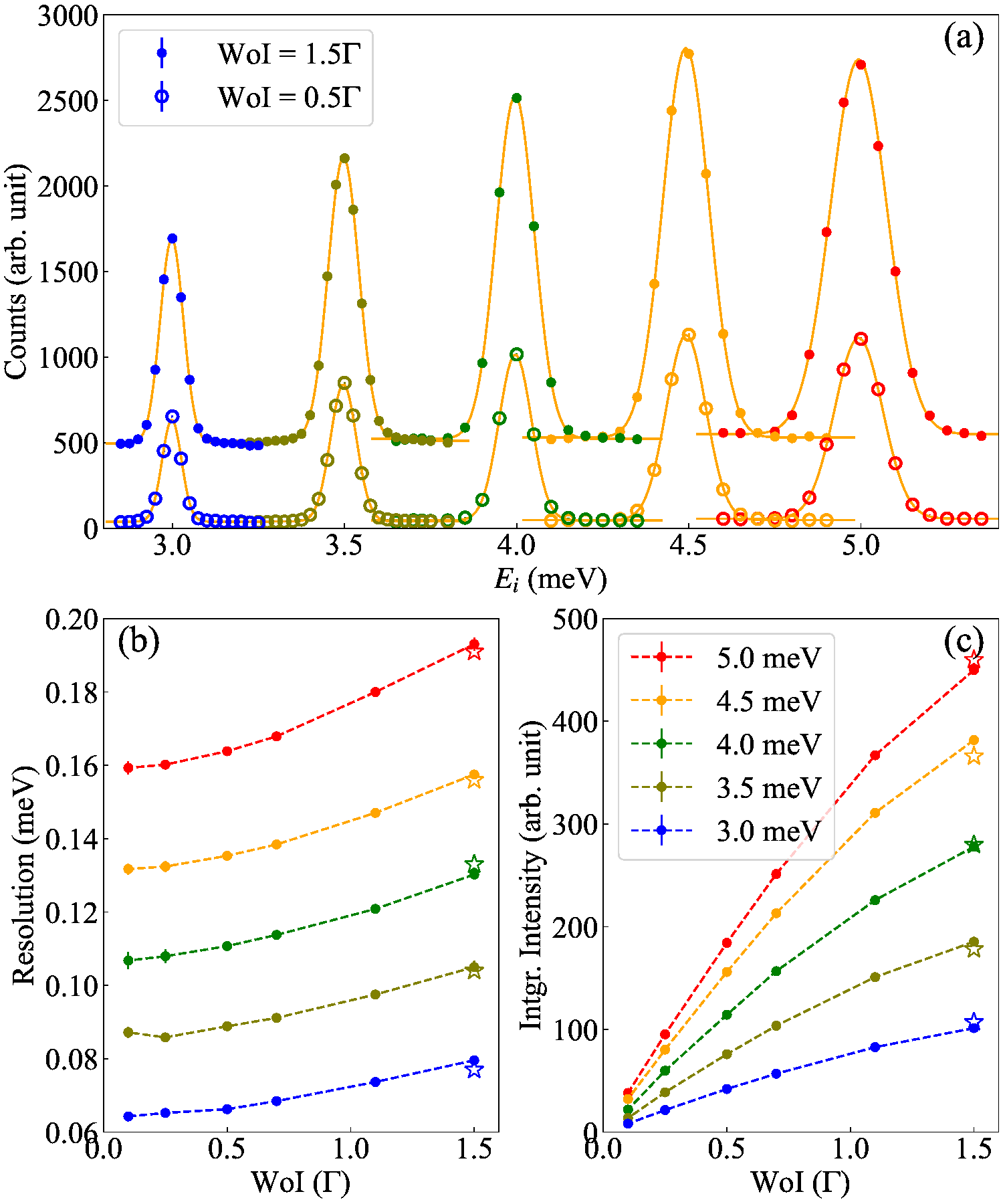}
		\caption{ (a) The intensity, summed over different widths of focused image peaks on PSDs (WoI) and averaged for all tubes, is plotted as a function of the scanning $E_i$. The solid and open symbols represent data of WoI = $1.5 \Gamma$ and $0.5 \Gamma$, respectively. The $1.5 \Gamma$ data is shifted upwards for clarity. 
		(b) Energy resolution, represented by the full widths at half maximum of peaks in (a), as a function of WoI. 
		(c) Integrated intensity as a function of WoI. 
		In all figures, data from different $E_f$ channels are color-coded as indicated in (c).
		The open star symbols in (b) denote the McStas simulated values and in (c) the expected values of $\Omega * V_\mathrm{res}/\mathrm{HON}$ (see text).
		}
		\label{eres}
	\end{figure}

	\section{Commissioning}		
			
	\begin{table*}[tbhp]
		\caption{The specifications of the DCRF analyzers for five fixed final energies. The parameters are organized and presented in three categories: focusing geometrical parameters, normalization parameters and experimental characterization results. The HOPG crystals have mosaic spreads of 0.4\textdegree{}--0.5\textdegree{}.}
		\label{table_ana}
		\begin{ruledtabular}
		\begin{tabular}{cccccc}
			$E_f$ (meV)	& 3.0 & 3.5 & 4.0 & 4.5 & 5.0 \\
			
			\hline
			& \multicolumn{5}{c}{Focusing Geometrical Parameters} \\
			$2\theta_A$ (deg.) & 102.24 & 92.23 & 84.78 & 78.93 & 74.17\\
			
			$L_\mathrm{S}$ (mm) & 960 & 1030 & 1105 & 1182 & 1265 \\
			
			$L_\mathrm{I}$ (mm) 
			\footnote{$L_\mathrm{I} = H/\sin2\theta_A$, where $H = 500\,\mathrm{mm}$, see Fig.~\ref{focus}(d).}
			& 511.6 & 500.4 & 502.1 & 509.5 & 519.7 \\
			
			$L_\mathrm{S}/L_\mathrm{I}$ & 1.9 & 2.0 & 2.2 & 2.3 & 2.4 \\
			
			$R_\mathrm{Rowland}$ (mm) & 505.2 & 564.2 & 629.9 & 700.0 & 775.9 \\
			
			$L^{\prime}_\mathrm{S}$ (mm) 
			\footnote{The distance from sample to the second column HOPG crystals, displaced by 38\,mm from the first column, i.\,e.,
				$L^{\prime}_\mathrm{S} = L_\mathrm{S} + 38$.} 
			& 998 & 1068 & 1143 & 1220 & 1303 \\
			
			$R^{\prime}_\mathrm{Rowland}$ (mm) 
			\footnote{The radius of the second Rowland circle.} 
			& 522.1 & 581.3 & 647.5 & 718.1 & 794.6 \\

			\hline
			& \multicolumn{5}{c}{Normalization Parameters} \\
			HOPG Width (mm) 
			\footnote{The effective width, subtracting a 2\,mm margin on each side, which is shielded for crystal mounts.}
			& 26 & 28 & 30 & 31 & 32 \\	
			
			Horizontal Coverage (deg.) & 1.55 & 1.56 & 1.56 & 1.50 & 1.45 \\
			
			HOPG Height (mm) 
			\footnote{The total height, summing for 7 blades and projected into the vertical direction.}
			& 65.4 & 65.6 & 70.1 & 69.9 & 71.8 \\
			
			Vertical Coverage (deg.) & 3.90 & 3.65 & 3.64 & 3.39 & 3.25 \\
			
			Solid Angle $\Omega$ (mrad.) & 1.84 & 1.73 & 1.72 & 1.55 & 1.44 \\
			$V_\mathrm{res}$ ($\mathrm{\AA^{-3}}$)
			\footnote{The resolution volume $V_\mathrm{res} = k_f^3/\tan\theta_A$.}
			& 1.40 & 2.11 & 2.94 & 3.89 & 4.96 \\
			
			HON
			\footnote{The correction factor for high-order neutrons in monitor counts, measured via the velocity selector on XINGZHI \cite{XINGZHI}.}
			& 2.65 & 2.25 & 1.98 & 1.81 & 1.70 \\
			
			$\Gamma$ (mm)\footnote{The Gaussian-fit FWHM of the focused double peaks, averaged for angular channels, explained in the text.}
			& 14.1, 13.4 & 12.4, 12.3 & 12.0, 12.4 & 12.0, 12.8 & 12.7, 13.9 \\
			\hline
			& \multicolumn{5}{c}{Experimental Results at WoI = $1.5 \Gamma$} \\

			$\Delta E^{\mathrm{FWHM}}_{\mathrm{exp}}$ (meV) 
			& 0.079 & 0.105 & 0.130 & 0.158 & 0.193 \\
			$\Delta E^{\mathrm{FWHM}}_{\mathrm{McStas}}$ (meV) 
			& 0.077 & 0.104 & 0.133 & 0.156 & 0.191 \\
			
			Ingtr. I (a.\,u.) & 101.2 & 185.3 & 277.9 & 381.7 & 449.9 \\
			
			$I*\mathrm{HON}/\Omega/V_\mathrm{res}$ (a.\,u.) 
			\footnote{Normalized for their average value.}
			& 0.945 & 1.040 & 0.993 & 1.044 & 0.978 \\
				
		\end{tabular}
		\end{ruledtabular}
	\end{table*}

	\begin{figure}[tbhp]
		\centering
		\includegraphics[width=\columnwidth]{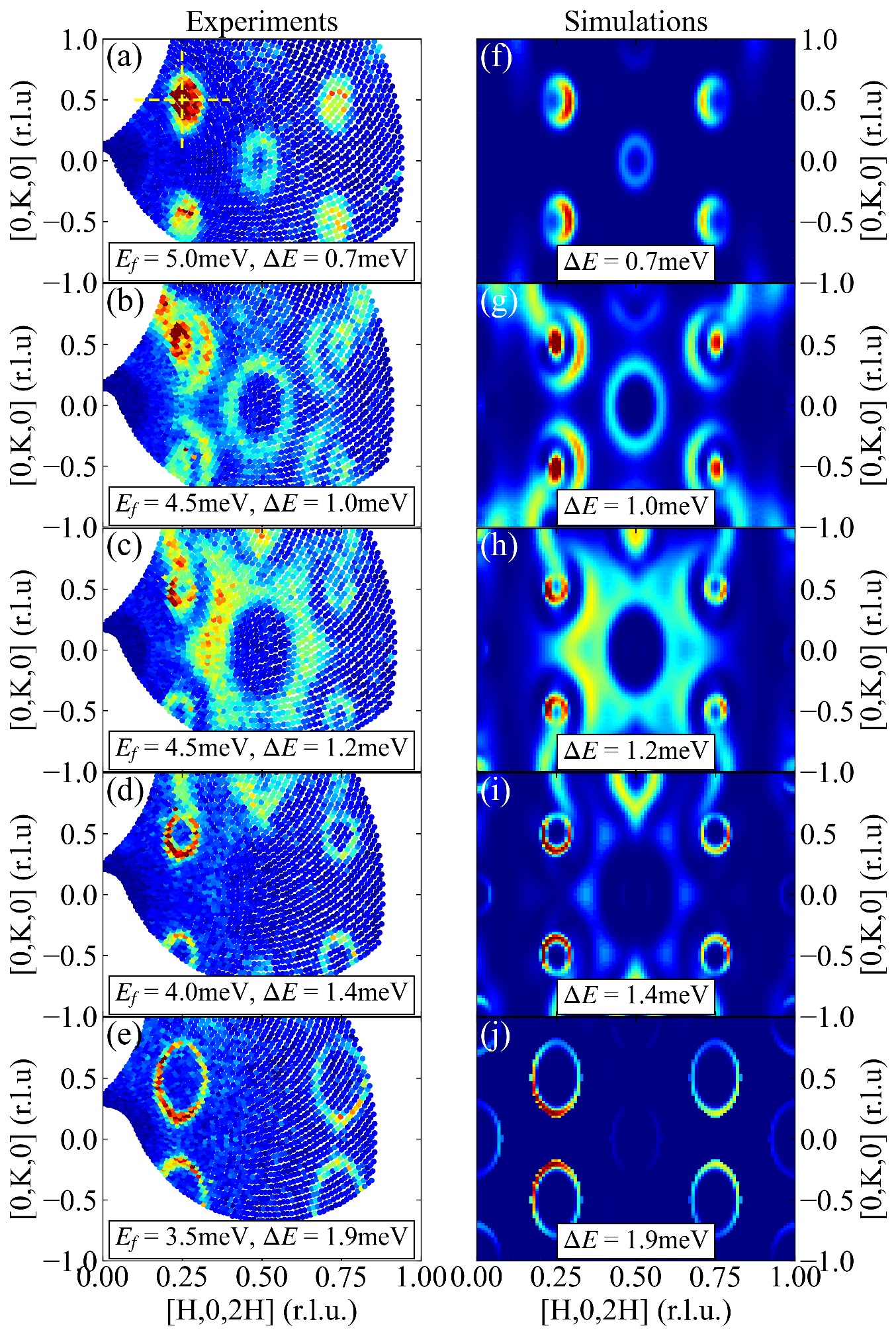}
		\caption{Excitation maps from experiments (a-e) and SpinW siumlations (f-j) at constant energy transfers of 0.7, 1.0, 1.2, 1.4, 1.9\,meV. In (a), the two directions [H,0.5,2H] and [0.25,K,0.5] are marked. 
		}
		\label{ecut}
	\end{figure}
	
	\begin{figure}[tbhp]
		\centering
		\includegraphics[width=\columnwidth]{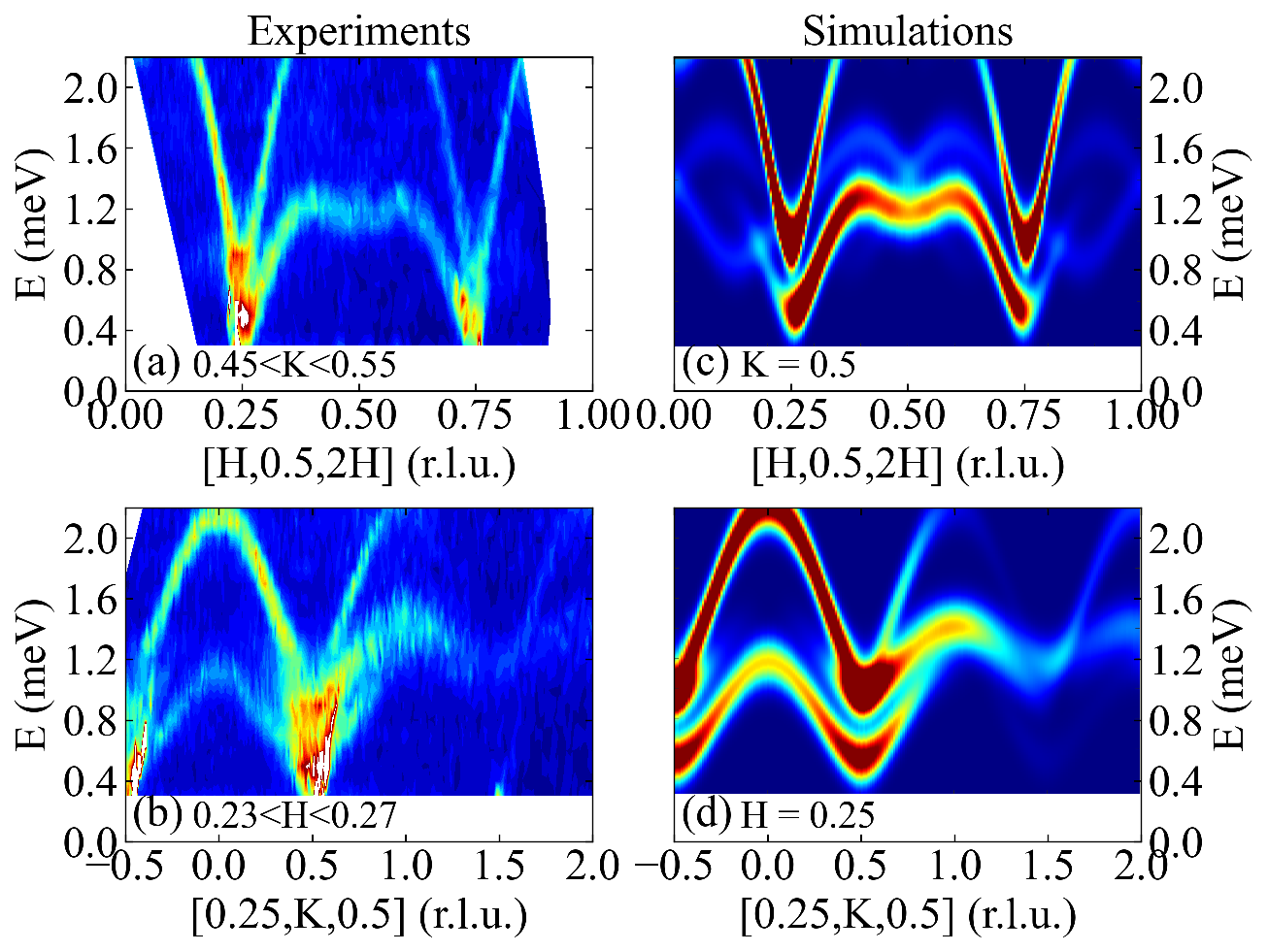}
		\caption{Spin wave dispersion cuts along the [H,0.5,2H] (a,c) and [0.25,K,0.5] (b,d) directions. 
		In (a), data is taken from the 0.5$\pm$0.05 range along the [0,K,0] direction, and in (b) from the 0.25$\pm$0.02 range along the [H,0,2H] direction. For simplicity, a constant energy resolution of 0.25\,meV was applied in the SpinW simulations presented in (c) and (d).
		}
		\label{disp}
	\end{figure}

	The instrument was commissioned at the CARR reactor operating at a reduced power of 30\,MW. With the cold source off, the monochromator delivered approximately $1\times10^6\,\mathrm{n/cm^2/s}$ neutrons at incident energy $E_i = 5\,\mathrm{meV}$ under vertical focusing. If at the full power of the reactor and with the cold source turning on, we expect a flux on sample of $5\times10^7\,\mathrm{n/cm^2/s}$ at $E_i = 5\,\mathrm{meV}$ \cite{XINGZHI}.
	The Be-filter after the sample was cooled to around 100\,K, and no collimations were used in the beam. 
	
	A vanadium pellet with a diameter and height of both 6.35\,mm was placed at the sample position.
	The focusing effect of the double-column analyzers was evaluated by measuring the incoherent elastic scattering of vanadium at fixed $E_i$ of 3.0, 3.5, 4.0, 4.5, 5.0\,meV.  
	In Fig.~\ref{detview}, data from these five $E_i$ channels are shown.
	Here, an average background of about 3 counts per minute per inelastic channel (integrated over 1.5 FWHM of the double-focused peaks on 3 tubes) was subtracted. No more background subtraction was performed in subsequent figures.
	Five pairs of focused peaks appear on each PSD tube (expect for the rightmost three channels that are blocked by the beam stop), consistent with our DCRF design.
	
	We present the intensity distribution along tube \#24 (the first PSD tube in angular channel \#8) in Fig.~\ref{detview}(b). Two distinct sharp peaks are visible at each energy, representing well focused images.
	The peaks at $E_i = 5\,\mathrm{meV}$ are shown in greater detail in Fig.~\ref{detview}(c), with blue shades, red dashed lines, and red shades	indicating the 0.5, 1 and 1.5 multiples of the full width at half maximum (FWHM, denoted by $\Gamma$) of the focused image peaks, respectively.
	All image peaks are fitted with Gaussian profiles in a consistent manner, with their centers and FWHM shown in Fig.~\ref{detview}(d). The uniform distribution of peaks across all tubes, aligned with the calculated positions, confirms the successful implementation. 	

	The centers and widths of the image peaks were stored in a file for subsequent data reduction. We developed a Python-based data reduction program that converts the neutron counts on PSD pixels into the intensity distribution as a function of momentum and energy transfer, normalizes the intensity, and visualizes the data in the needed direction. Details of the program will be described elsewhere.
	
	The intensity of each $E_f$ channel is obtained by summing neutron counts on PSD pixels where the double image peaks locate.
	The summation range, called width of interest (WoI) in the unit of the measured FWHM ($\Gamma$) thereafter, would affect the ultimate resolution and intensity.	
	These were evaluated as $E_i$ was scanned from 2.85 to 5.35\,meV.
	Presented in Fig.~\ref{eres}(a), the curves with neutron counts summing from WoI = $0.5 \Gamma$ and WoI = $1.5 \Gamma$ are compared. It is evident that the data of $0.5 \Gamma$ is sharper in energy but weaker in intensity.
	The energy resolution is defined as the peak width in the $E_i$-scanning curve ($\Delta E^{\mathrm{FWHM}}_{\mathrm{exp}}$).  
	The changes in energy resolution and integrated intensity by varying the summation window WoI are presented in Fig.~\ref{eres}(b,c). 
	The intensity decreases more dramatically than the resolution improves.
	Reducing from WoI = $1.5 \Gamma$ to $0.5 \Gamma$ results in a roughly 2.4-fold drop in intensity and 15\% resolution improvements.
	When the WoI is reduced below $0.5 \Gamma$, the intensity continues to decrease, but the resolution barely changes.
	These observations guide the selection of WoI for future experiments. 
	The best resolution is achieved at $E_f = 3\,\mathrm{meV}$, measuring approximately 0.066\,meV at WoI = $0.5 \Gamma$, or 0.079\,meV at WoI = $1.5 \Gamma$. Even higher resolutions can be expected via adding collimations in the beam at sacrifice of intensity.

	The specifications and characterizations of the DCRF multiplexing backend are listed in Table~\ref{table_ana}. We can see that the energy resolution $\Delta E^{\mathrm{FWHM}}_{\mathrm{exp}}$ measured at WoI = $1.5 \Gamma$, closely matches the McStas simulated value of $\Delta E^{\mathrm{FWHM}}_{\mathrm{McStas}}$ within 3\%.
	Additionally, the integrated intensities at different $E_f$, after corrected for high-order neutrons in the monitor counts ($I*\mathrm{HON}$), align well with the expected values of $\Omega * V_\mathrm{res}$ for vanadium incoherent scattering, where $\Omega$ is the solid angle of analyzer and $V_\mathrm{res}$ is the resolution volume \cite{TripleAxis,MultiFLEXX3}. 
	The consistency between commissioning results and expected values, in both resolution and intensity, is also demonstrated in Fig.~\ref{eres}(b,c).
	The results further confirm the successful design and implementation of DCRF analyzers.

	To demonstrate the capability of BOYA, we performed inelastic neutron scattering measurements on MnWO$_4$, a multiferroic material whose spin wave excitations have been comprehensively studied \cite{MnWO4_Ye,MnWO4_Xiao,MnWO4_Wang}. 
	Below 7\,K, it orders into the AF1 magnetic phase ($\uparrow \uparrow \downarrow \downarrow$ configuration) with the wave-vector of (0.25,0.5,0.5).
	A single crystal measuring approximately 5 $\times$ 5 $\times$ 10\,$\mathrm{mm}^3$ in size and 1.6\,g in mass was aligned into the (H,K,2H) plane and cooled to 4\,K. Reciprocal lattice units (r.l.u.) for $\boldsymbol{Q}$ were used to describe the results.
	Four MADS positions were used to cover the 7\textdegree{} gap between angular channels. At each MADS position, the sample was rotated by 100\textdegree{} at a rate of 2\textdegree{} per 5\,min. This took about 16.6 hours for one incident energy at the reduced CARR capacity. Data were collected at eight different $E_i$ from 5.0 to 5.7\,meV in 0.1\,meV steps. This resulted in about 40\% redundancy, with some energy transfers being double-collected for commissioning purpose (e.\,g. $E_i = 5.6\,\mathrm{meV}$, $E_f = 5\,\mathrm{meV}$ v.s. $E_i = 5.1\,\mathrm{meV}$, $E_f = 4.5\,\mathrm{meV}$).
	In total it took about 5.5 days of beamtime to obtained a 3D data set for the spin wave excitation, with the cold source turned off. If the cold source can be activated, we expect to get the excitation map in half a day.
	
	The maps of constant energy excitation and dispersion were obtained using our python-based data analysis package. 
	In Fig.~\ref{ecut}, the spin wave signal at constant energy matches nicely with simulations via SpinW package\cite{spinw} using published model parameters\cite{MnWO4_Wang}. 
	The dispersion along the [H,0.5,2H] (or [0.25,K,0.5]) direction, obtained by binning data in the $0.45<\mathrm{K}<0.55$ (or $0.23<\mathrm{H}<0.27$) range, and stacking data of different energy transfers, is shown in Fig.~\ref{disp}. Spin wave branches are clearly observed and consistent with simulations.

	\section{Conclusion and Outlook}
	A multiplexing cold neutron spectrometer BOYA has been designed and constructed at CARR. Innovative double-column Rowland focusing analyzers have been employed to optimize both intensity and energy resolution. 
	It yields about 2-fold intensity gain compared with the traditional Rowland focusing.
	Commissioning results show well focused spots on all detectors at each corresponding energy, demonstrating the success of the design. The intensity and energy resolution obtained from the incoherent elastic scattering of the vanadium sample agree with the simulated values. Spin wave excitation on MnWO$_4$ was obtained efficiently in the 3D $(k_x,k_y,E)$ phase volume which are consistent with simulations.
	
	There is still space for further improvements. Firstly, the signal-to-noise ratio can be enhanced through better shielding. Specifically, the aluminum analyzer supports (Fig.~\ref{ana_real}) are not shielded. We expect coating them with neutron absorbing materials could lower the background. 
	Additionally, the MADS tank is not pumped, evacuating the tank could further reduce the background. 
	Importantly, it is estimated that at least an order of magnitude flux gain for cold neutrons can be achieved by turning on the cold source. This would significantly improve the signal noise ratio, reducing the nowadays experimental time from 3-7 days to less than one day. 
	
	Equipped with various sample environments including low temperature, magnetic field, and high pressure,
	BOYA has already contributed to scientific research in areas such as magnetic excitations \cite{CrOOD}, phonons\cite{Cs3Bi2Br9,song_BaSrNiAs}, and diffuse scattering\cite{TbScO3} since 2020.
	The double-column Rowland focusing design will contribute to the ongoing development of multiplexing neutron spectrometers and benefit the wide neutron scattering community.
	
	\section*{Acknowledgement}
	The authors gratefully acknowledge stimulating discussions with 
	C.~Broholm from Johns Hopkins University, 
	P.~B\"{o}ni from Technical University of Munich and SwissNeutronics,
	P.~Link from Heinz Maier-Leibnitz Zentrum and Technical University of Munich, 
	R.~Robinson from Australian Nuclear Science and Technology Organisation (now in University of Wollongong and Ibaraki University),
	J.~Lynn and Yiming Qiu from NIST Center for Neutron Research, 
	S.~Raymond from University Grenoble Alpes, 
	A.~Schneidewind from the J\"{u}lich Center for Neutron Science, 
	K.~Habicht from Helmholtz-Zentrum Berlin, 
	and F.~Groitl from Paul Scherrer Institute.
	The authors are indebted to Junfeng Li, Wenwei Shen, Lijie Hao, Hongliang Wang, Jianfei Qin for their contributions in engineering implementation. 
	Special thanks to Prof.~Yinguo Xiao from Peking University Shenzhen Graduate School for providing the high quality MnWO$_4$ crystals in our commissioning. 
	This work is supported by the National Key R{\&}D Program of China (Grant No.~2023YFA1406500), NSFC Grants (No.~12004426, No.~U2030106, No.~12304185) and Special Fund for Research on National Major Research Instruments of NSFC (No.~11227906).
	

	\section*{DATA AVAILABILITY}
	The data that support the findings of this study are available
	from the corresponding author upon reasonable request.
	
	\bibliography{boya.bib}{}
	
\end{document}